\documentclass[fleqn,usenatbib]{mnras}

\usepackage{newtxtext,newtxmath}

\usepackage[T1]{fontenc}

\DeclareRobustCommand{\VAN}[3]{#2}
\let\VANthebibliography\thebibliography
\def\thebibliography{\DeclareRobustCommand{\VAN}[3]{##3}\VANthebibliography}

\usepackage{graphicx}	
\usepackage{amsmath}	

\usepackage{academicons} 
\newcommand{\orcid}[1]{\href{https://orcid.org/#1}{\includegraphics[width=8pt]{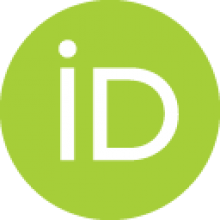}}}
\usepackage{CJK}

\title[Constrain CCC+TL cosmology with $H(z)$]{Stringent constraint on the CCC+TL cosmology with $H(z)$ Measurements}

\author[Lei et al.]{
Lei Lei (雷磊)$^{1,2}$ \orcid{0000-0003-4631-1915},
Ze-Fan Wang  (王泽凡)$^{1,2}$ \orcid{0009-0004-9366-1947},
Tong-Lin Wang  (王彤琳)$^{1,2}$  \orcid{0009-0004-0029-6080},
Yi-Ying Wang (王艺颖)$^{1}$  \orcid{0000-0003-1215-6443},
\newauthor{Guan-Wen Yuan (袁官文)}$^{3,4}$  \orcid{0000-0002-4538-8526},
Wei-Long Lin (林炜龙)$^{1,2}$ ,
Yi-Zhong Fan (范一中)$^{1,2}$ \orcid{0000-0002-8966-6911} \thanks{E-mail: yzfan@pmo.ac.cn (Y.Z. Fan)}
\\
$^{1}$Key Laboratory of Dark Matter and Space Astronomy, Purple Mountain Observatory, Chinese Academy of Sciences, Nanjing 210023, China\\
$^{2}$School of Astronomy and Space Science, University of Science and Technology of China, Hefei 230026, China\\
$^{3}$Department of Physics, University of Trento, Via Sommarive 14, 38123 Povo (TN), Italy\\
$^{4}$Trento Institute for Fundamental Physics and Applications (TIFPA)-INFN, Via Sommarive 14, 38123 Povo (TN), Italy
}

\date{Accepted 2026 February 26. Received 2026 February 26; in original form 2025 November 12}

\pubyear{\the\year{}}

\begin{document}
\begin{CJK*}{UTF8}{gbsn}
\label{firstpage}
\pagerange{\pageref{firstpage}--\pageref{lastpage}}
\maketitle

\begin{abstract}
Recently, the Covarying Coupling Constants and Tired Light (CCC+TL) hybrid model was proposed to explain the unexpectedly small angular diameters of high-redshift galaxies observed by the James Webb Space Telescope (JWST) that are challenging to reconcile with the $\Lambda$CDM model. In this work, we test the CCC+TL model against model-independent Hubble parameter [$H(z)$] measurements obtained from cosmic chronometers. It turns out that the parameter set optimized for the type-Ia supernova (SN Ia) dataset within the CCC+TL model fails to reproduce the $H(z)$ data, but the $\Lambda$CDM model works well. Statistical comparison using the $\Delta\chi^2$ strongly favors $\Lambda$CDM over CCC+TL for the $H(z)$ data, with $\Delta \chi^2 = 61.52$. Crucially, the CCC+TL framework exhibits a severe internal tension, where the SN Ia-optimized speed-of-light variation index $\alpha$ is rejected by the $H(z)$ dataset with a likelihood ratio of $\mathcal{R} \approx 1.7 \times 10^{-14}$. Our result suggests that the tension posed by JWST observations of compact high-$z$ galaxies may originate from the intrinsic properties and evolution of galaxies in the early universe.

\end{abstract}




\section{Introduction}
\label{sec:intro}

The launch of the James Webb Space Telescope (JWST) in late 2021 heralded a revolutionary era in observational cosmology \citep{2024PhRvL.132f1002S,2025PhRvL.134g1003R,2025PhRvL.134g1002J,2024Natur.628..277G,2024ApJ...975..285W,2025MNRAS.538.1264C,2024RAA....24l5012W,2025arXiv250619589L,2023PhRvD.107d3502H}. Its unprecedented infrared sensitivity and angular resolution have pierced the veil of the early universe, revealing galaxies at high redshifts and cosmic epochs previously inaccessible \citep{2023ApJ...946L..13F,2023Natur.622..707A,2023NatAs...7..611R,2024SCPMA..6729811L}. 

While confirming the immense power of modern astronomy, these observations have simultaneously posed profound challenges to the established cosmological paradigm \citep{2023Natur.616..266L,2023NatAs...7..731B,2024Natur.635..311X,2023ApJ...954L..48W,2023MNRAS.524.3385G,2025MNRAS.542L..19W,2024JCAP...05..097F,2024RAA....24d5001W,2024PDU....4601587M,2024ApJ...967..172S,2024ApJ...970...63L,2025MNRAS.538.3210B}, the $\Lambda$ Cold Dark Matter ($\Lambda$CDM) model. In these studies, some suggested that the observed star-forming efficiency is too high or that the mass of galaxies exceeds original expectations \citep{2023Natur.616..266L,2023NatAs...7..731B,2024Natur.635..311X}.
To better understand these high-redshift galaxies, numerous models have been suggested. These include the rapid formation of massive population III stars or dark stars \citep{2023PNAS..12005762I,2024MNRAS.527.5929Y,2024PDU....4401496I,2025ApJ...980..249L},the existence of dark matter and dark energy~\citep{2023JCAP...10..072A,2024RAA....24a5009L, 2024JCAP...07..072M}, cosmic string \citep{2023SCPMA..6620403W,2023PhRvD.108d3510J}, feedback-free processes \citep{2023MNRAS.523.3201D,2025ApJ...988L..35W,2024A&A...690A.108L},  the presence of primordial black hole seeds~\citep{2024SCPMA..6709512Y} and alterations in the power spectrum \citep{2023ApJ...953L...4P,2024PhRvL.132f1002S}. In addition to the challenges associated with galaxy formation, there are also the age of galaxies \citep{2024ApJ...970...63L,2024ApJ...967..172S} and the apparent diameter of galaxies \citep{2022Galax..10..108L,2023MNRAS.524.3385G,2010IJMPD..19..245L}. In order to solve the problem that the apparent diameter of galaxies observed by JWST is lower than the prediction of $\Lambda$CDM, a noteworthy one involves applying the tired light concept \citep{2022Galax..10..108L}, extending the Universe's age to allow sufficient time for high-redshift galaxies to form. This concept is developed by ~\citet{2023MNRAS.524.3385G}, named ``Covarying Coupling Constants and Tired Light (CCC+TL)  hybrid model''. 

In this work, we undertake an evaluation of the CCC+TL hybrid model. We begin by providing a concise introduction to the CCC and TL hybrid model. Subsequently, we employ the best-fit parameters derived from the Pantheon+ SNe Ia sample to evaluate if the defined parameter vector yields $H(z)$ values align with CC measurements. In addition, we conduct a free fitting procedure on the CCC+TL model with the $H(z)$ dataset to examine whether the resulting posterior distribution of the parameters is consistent with the favored parameters of the SN Ia dataset. We find that the CCC+TL hybrid model faces significant challenges.

\section{CCC+TL Hybrid Model}\label{sec:model}

The CCC+TL hybrid model departs from  $\Lambda$CDM by invoking two core mechanisms designed to eliminate the need for both dark energy and dark matter. First, the model assumes a varying speed of light ($c$), where evolves covariantly with the scale factor of the Universe. This variation modified the relationship between redshift, time, and distance, leading to distinctive predictions for cosmological observations \citep{2020MNRAS.498.4481G,2022MPLA...3750155G,2022MNRAS.511.4238G,2024Univ...10..266G}. Second, it incorporates a reinterpretation of the tired light (TL) hypothesis \citep{1929PNAS...15..773Z}, where photons gradually lose energy as they propagate through space, contributing to the observed redshift \citep{2024Parti...7..703S,2018IJAA....8..219G}. The combination of these mechanisms increases the angular diameter distance $D_A$ at a given redshift, thereby offering a naturally explanation for the unexpectedly small angular diameters of high-$z$ JWST galaxies (note that $\theta(z) = D_{\mathrm{intrinsic}} / D_A$). Moreover, the age of the Universe in this framework is significantly extended to $t \approx 26.7\, \rm Gyr$, significantly larger than $\Lambda$CDM's estimate, alleviates the problem of excessive early mass growth in galaxies and black holes inferred from JWST data. 

The CCC+TL model has also been shown to provide a statistically acceptable fit to the Pantheon+ cosmological distance modulus of SNe Ia \citep{2023MNRAS.524.3385G}, reproducing the observed Hubble diagram up to $z \approx 2.3$. Furthermore, the CCC+TL model is validated by baryon acoustic oscillation (BAO) measurements, as claimed in \citet{2024ApJ...964...55G}.

Despite these progress, the CCC+TL model should be scrutinized against independent cosmological probes not reliant on SNe Ia or galaxy size indicators. Moreover, any viable alternative to $\Lambda$CDM must reproduce both the SNe Ia Hubble diagram and the $H(z)$ dataset with a single, self-consistent parameter set. Here we focus on the Hubble parameter ($H(z)$) measured from cosmic chronometers (CC) -- passively evolving galaxies. The method, originally proposed by \citet{2002ApJ...573...37J}, utilizes differential ages of these CC galaxies to yield $H(z) = -(1+z)^{-1} dz/dt$. As the methodology has evolved, CC measurements now encompass different techniques, including the $D_n4000$ break, Lick indices, and full spectrum fitting \citep{2024arXiv241201994M,2023ApJS..265...48J,2023JCAP...11..047J,2022ApJ...928L...4B,2023A&A...679A..96T,2025MNRAS.540.3135L,2026arXiv260107345W}. Crucially, these techniques uniquely utilize stellar population ages and spectral data, eliminating dependencies on cosmological models, standard-candle mechanisms (like SNe Ia), or universe geometry postulates \citep{2010AdAst2010E..81Z,2020ApJ...898...82M}.
Consistency between a model's prediction for $H(z)$ and CC measurements thus serves as a critical test of its viability \citep{2023MNRAS.523.3406F,2024MNRAS.527.4874C,2025arXiv250211625N,2025MNRAS.542.1063H,2026MNRAS.546ag303F}. In the following sections, we assess whether the CCC+TL model meets this critical requirement.

\section{Data and Analysis}
\label{sec:data}
To independently evaluate the CCC+TL hybrid model, we adopt the same formulations as presented in \citet{2023MNRAS.524.3385G}. In this framework, the Hubble parameter is expressed as
\begin{equation}
    H = \left(H_{0,\mathrm{CCC}}+\alpha\right)\left(1+z_{\mathrm{CCC}}\right) \left[f^{-\frac{1}{2}} \left(1+z_{\mathrm{CCC}}\right)^{\frac{1}{2}}+\frac{3H_{0,\mathrm{CCC}}+\alpha}{2H_{0,\mathrm{CCC}}}\right] - \alpha.
    \label{eq:H_z}
 \end{equation}
The redshift evolution function $f(z_{\mathrm{CCC}})$, encoding the expansion history due to the covariation of the speed of light and gravitational constant, is given by 
\begin{equation}
    f^{-1 / 2} = \left( -\frac{D}{2 A} + B^{1 / 2} \right)^{1 / 3} + \left( -\frac{D}{2 A} - B^{1 / 2} \right)^{1 / 3},
    \label{eq:f_z}
 \end{equation}
where $A = 1 - \frac{3}{2}\frac{\left( H_{0,\mathrm{CCC}} + \alpha \right)}{\alpha} = 1 - C$, $D = -(1 + z_{\mathrm{CCC}})^{-3/2}$, and $B = \left( -\frac{D}{2 A} \right)^2 + \left( \frac{C}{3 A} \right)^3$.

\begin{figure}
\centering
 \includegraphics[width=\linewidth]{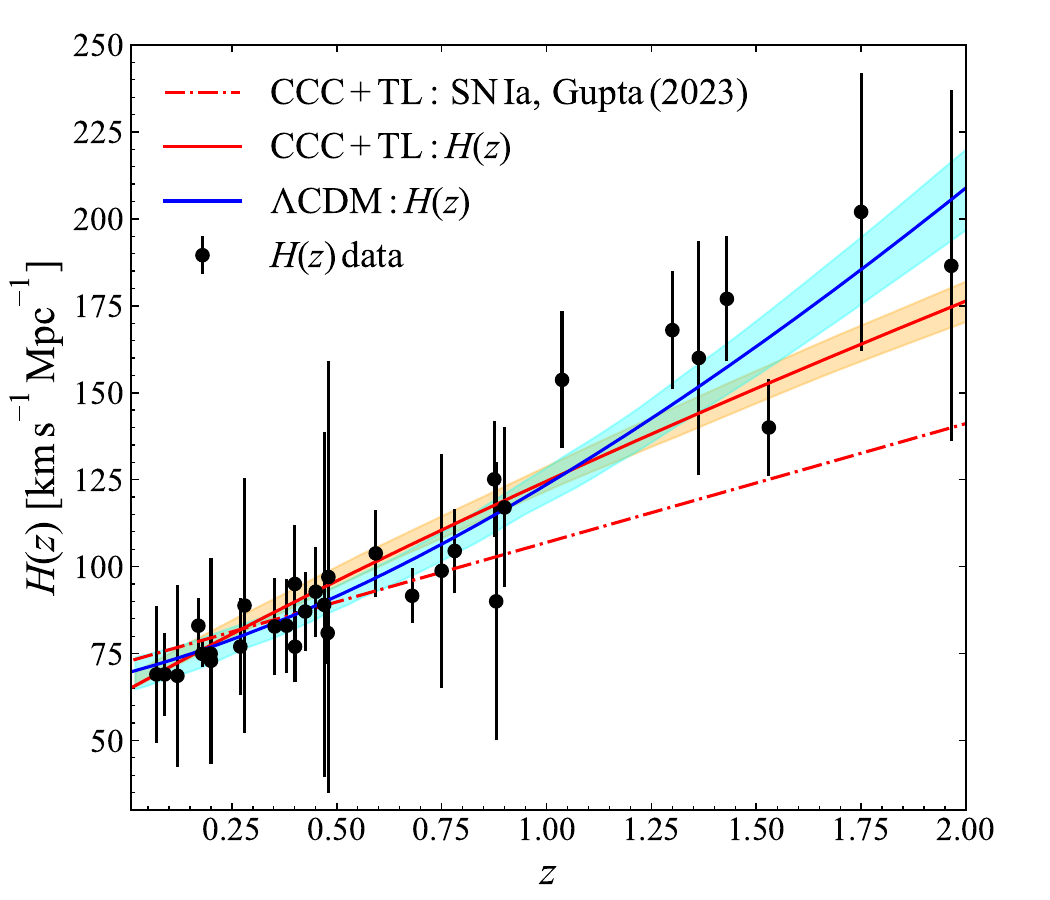}
 \caption{The red and blue solid lines show the best fits of the CCC+TL and $\Lambda$CDM models, respectively, to the $H(z)$ data. The color regions represent the $1\sigma$ uncertainties.  The red dash-dotted line is the  ``predicted $H(z)$'' with the CCC+TL best-fit parameters of \citet{2023MNRAS.524.3385G}, which significantly deviates from the observation data (the black error bar points).} 
 \label{fig:data}
\end{figure}

To confront this model with observation, we use 32 measurements of the Hubble parameter $H(z)$ obtained via the CC method, covering the redshift range $0.01 < z < 2$ \citep{2024arXiv241201994M}. We adopt these 32 independent $H(z)$ data points \footnote{organized at: \url{https://apps.difa.unibo.it/files/people/Str957-cluster/astro/CC_data/data_CC.dat}} to test CCC+TL cosmology \citep{2014RAA....14.1221Z,2005PhRvD..71l3001S,2012JCAP...08..006M,2015MNRAS.450L..16M,2016JCAP...05..014M,2010JCAP...02..008S,2017MNRAS.467.3239R}. 
The data are displayed in Figure \ref{fig:data}.

It is crucial to note that the cosmic expansion history at high redshift is governed solely by the CCC component \citep{2023MNRAS.524.3385G}. Although the TL model contributes to the observed redshift, it does not contribute to the cosmological expansion history \citep{2023MNRAS.524.3385G}. Therefore, the redshift argument $z_{\mathrm{CCC}}$ in Eq. (\ref{eq:H_z}) specifically refers to the redshift component attributable to the CCC model.

The relationship between the total observed redshift ($z_{\mathrm{CCC+TL}}$) and the CCC redshift ($z_{\mathrm{CCC}}$) is given by combining Eqs. (44) from \cite{2023MNRAS.524.3385G}, i.e., 
\begin{equation}
    1 + z_{\mathrm{CCC+TL}} = \left(1 + z_{\mathrm{CCC}} \right) e^{y_{\mathrm{CCC}}},
    \label{eq:z_tot}
 \end{equation}
\begin{equation}
    y_{\mathrm{CCC}} =  I(z_{\mathrm{CCC}}) \left[ \frac{\left(H_{0,\mathrm{CCC}} + \alpha\right)}{2} \left( 3 + \frac{\alpha}{H_{0,\mathrm{CCC}}} \right) \right] ,
    \label{eq:y}
\end{equation}
\begin{equation}
 I(z_{\mathrm{CCC}}) = \int^{z_{\mathrm{CCC}}}_{0} \frac{dz'}{\left(H_{0,\mathrm{CCC}} + \alpha \right) (1+z')^{\frac{3}{2}}f(z')^{-\frac{1}{2}} - \alpha}. 
 \label{eq:I}
\end{equation}
Equation (\ref{eq:z_tot}) expresses the total observed redshift $z_{\mathrm{CCC+TL}}$. Given the total redshift and the CCC parameters, $z_{\mathrm{CCC}}$ can be computed numerically via interpolation using Eq. (\ref{eq:z_tot}). Since Eq. (3) is transcendental, we invert it numerically. We generate a dense logarithmic grid of $z_{\mathrm{CCC}}$ values ranging from $0.01$ to $10$. For each grid point, the integral $I(z_{\mathrm{CCC}})$ in Eq. (\ref{eq:z_tot}) is computed numerically to obtain the corresponding $z_{\mathrm{CCC+TL}}$. Since the mapping $z_{\mathrm{CCC}} \to z_{\mathrm{CCC+TL}}$ is strictly monotonic, we invert this relation using cubic spline interpolation to determine $z_{\mathrm{CCC}}$ for any measured $H(z)$ point. All numerical operations, including integration and interpolation, were performed using standard routines from the \texttt{NumPy} \citep{2020Natur.585..357H} and \texttt{SciPy} \citep{2020NatMe..17..261V} libraries. 

The TL redshift component is then obtained from
\begin{equation}
 z_{\mathrm{TL}} = \frac{z_{\mathrm{CCC+TL}} - z_{\mathrm{CCC}}}{1 + z_{\mathrm{CCC}}}.
 \label{eq:z_TL}
\end{equation}

\begin{figure*}
\centering
 \includegraphics[width=0.45\linewidth]{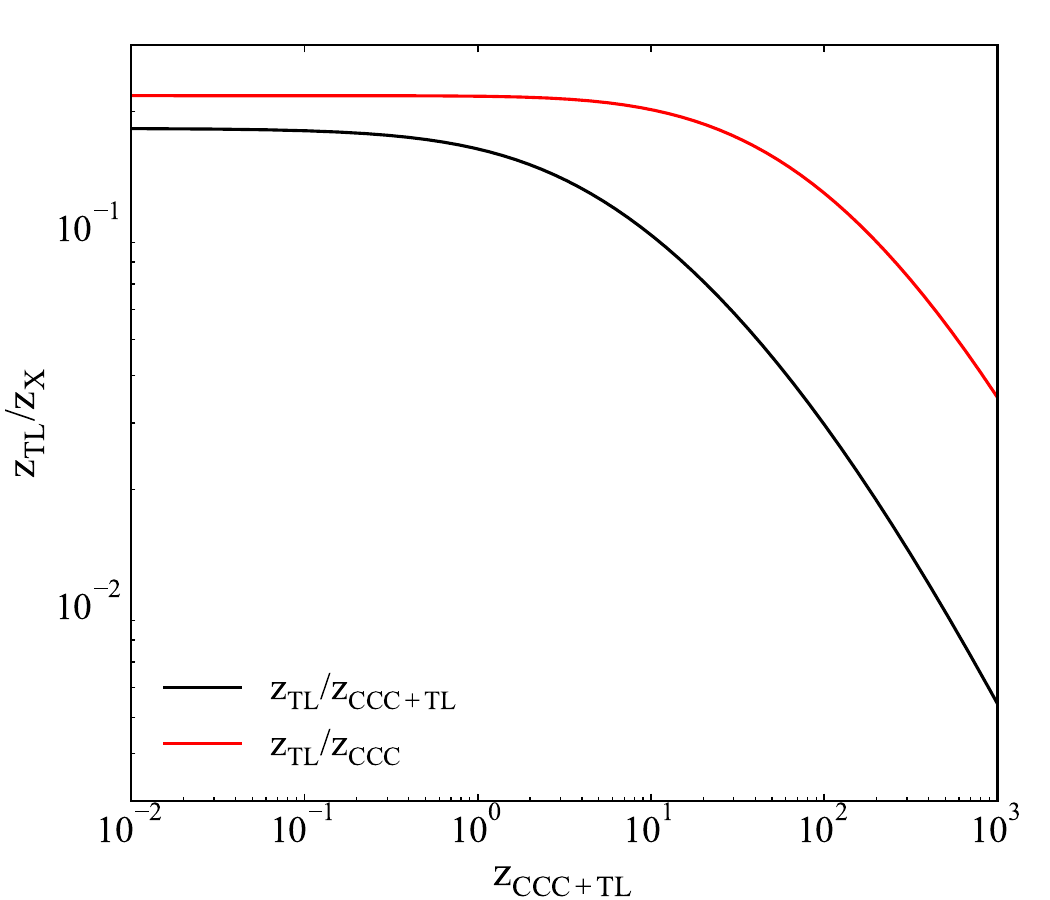}
 \includegraphics[width=0.45\linewidth]{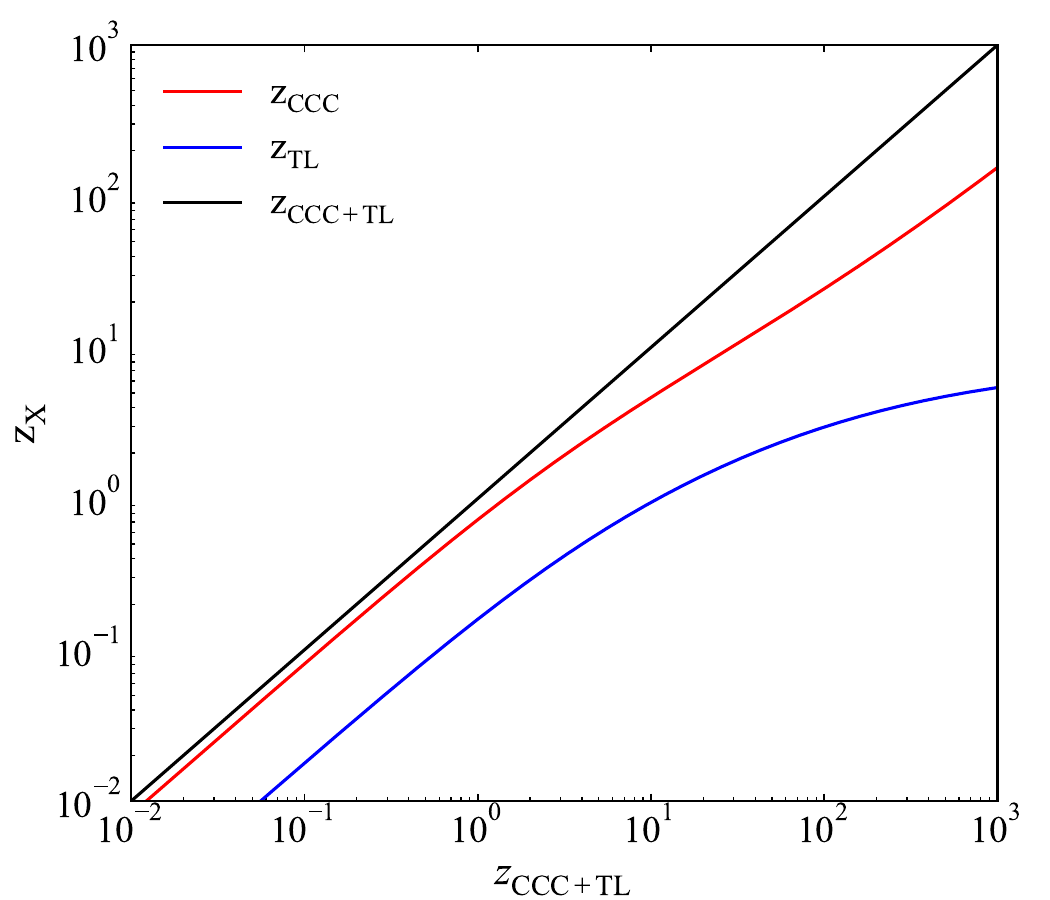}
 \caption{Redshift contributions of the CCC and TL components within the CCC+TL model, calculated using the best-fit SNe Ia parameters. Left Panel: Ratio of the TL redshift contribution to the CCC contribution (red) and to the total redshift $z_{\rm{CCC+TL}}$ (black). Right Panel: Proportion of the total redshift $z_{\rm{CCC+TL}}$ attributed to the CCC (red) and TL (blue) components.} 
 \label{fig:z_X}
\end{figure*}

Using Eqs. (\ref{eq:z_tot}-\ref{eq:z_TL}), we numerically compute the redshift contributions of CCC and TL across different redshifts for the CCC+TL cosmology. Figure~\ref{fig:z_X} shows these contributions calculated using the best-fit SNe Ia parameters from \cite{2023MNRAS.524.3385G}. Our results are consistent with \cite{2023MNRAS.524.3385G} for identical parameter values.

For comparison, we take the standard $\Lambda$CDM cosmological model. The Hubble parameter in $\Lambda$CDM is given by
\begin{equation}
    H_{\Lambda \mathrm{CDM}}(z) = H_0 \sqrt{\Omega_{0,\mathrm{m}} \left( 1 + z \right)^3 + 1 - \Omega_{0,\mathrm{m}}}.
    \label{eq:H_z_LCDM}
 \end{equation}

We perform a Bayesian analysis to estimate the posterior probability distributions of the model parameters. While the CC method yields cosmology-independent measurements of $H(z)$, it is subject to systematic uncertainties arising from factors such as metallicity, contamination by young stellar populations, and stellar population synthesis (SPS) models \citep{2020ApJ...898...82M,2022LRR....25....6M,2024arXiv241201994M}. Therefore, it is crucial to incorporate the total error budget—including these systematics—into the fitting procedure. 

Our dataset consists of 32 $H(z)$ measurements. For the subset derived from \citet{2012JCAP...08..006M,2015MNRAS.450L..16M,2016JCAP...05..014M,2017MNRAS.467.3239R}, systematic errors due to star formation history (SFH), young stellar components, and metallicity have been explicitly quantified. Furthermore, to account for correlations between data points, we utilize the full covariance matrix where available. Specifically, for 15 of these points, we adopt the covariance matrix $\mathbf{C}_{\mathrm{sub}}$ calculated by \citealt{2020ApJ...898...82M,2022LRR....25....6M,2024arXiv241201994M}\footnote{The covariance matrix is available at: \url{https://gitlab.com/mmoresco/CCcovariance}}. For the remaining 17 measurements where covariance estimates are unavailable, we treat them as uncorrelated (i.e., using only diagonal variances).

The log-likelihood function, assuming a Gaussian distribution, is given by:
\begin{equation}
\begin{aligned}
\ln \mathcal{L} = &-\frac{1}{2} \sum_{i=1}^{17} \left[ \frac{(H_{\mathrm{obs},i} - H_{\mathrm{model}}(z_{i},\Theta))^2}{\sigma_i^2} + \ln(2\pi \sigma_i^2) \right] \\
&- \frac{1}{2} \left[ \Delta \mathbf{H}_{15}^T \mathbf{C}_{\mathrm{sub}}^{-1} \Delta \mathbf{H}_{15} + \ln(\det(\mathbf{C}_{\mathrm{sub}})) + 15 \ln(2\pi) \right],
\end{aligned}
\label{eq:likeli}
\end{equation}
where $H_{\mathrm{obs},i}$ represents the observed Hubble parameter at redshift $z_i$, and $\sigma_{i}$ denotes the uncertainty for the 17 uncorrelated measurements. For the correlated subset, $\Delta \mathbf{H}_{15} = \mathbf{H}_{\mathrm{obs}} - \mathbf{H}_{\mathrm{model}}(\mathbf{z},\Theta)$ is the residual vector of the 15 measurements, and $\mathbf{C}_{\mathrm{sub}}$ is the corresponding covariance matrix. We sample the posterior distribution using the \textit{emcee} package \citep{2013PASP..125..306F}.

To fit the $H(z)$ data with the CCC+TL model, the free parameters include the local Hubble parameters $H_{0,\mathrm{CCC}}$ and the speed-of-light variation exponent $\alpha$. Table~\ref{tab:prior+post} lists the flat prior distributions used for these parameters. Due to potential inefficiency in constraining the low tail of $H_{0,\mathrm{CCC}}$ in linear space, we assumed a logarithmic prior for this parameter. For $\Lambda$CDM, we also used flat priors within the ranges specified in Table~\ref{tab:prior+post}. We adopt a uniform prior for $\alpha$ and a log-uniform prior for $H_{\rm 0,CCC}$ (see Table \ref{tab:prior+post}). The log-uniform prior on $H_{\rm 0,CCC}$ is chosen as a standard non-informative prior for a scale parameter. For $\alpha$, the wide prior range is selected to ensure it fully encompasses the parameter space preferred by SN Ia data ($\alpha = -47.59$; \citealt{2023MNRAS.524.3385G}), including its long negative tail, as well as the physically plausible region for $H(z)$. To ensure our results are not prior-driven, we performed a sensitivity test with a narrower prior range of $\alpha \in [-100, 100]$ and found the resulting posterior distributions to be robust and virtually identical to our fiducial run.

\begin{table}
\centering
    \caption{Prior ranges and posterior summaries (median and $1\sigma$ credible interval) for parameters fitted to the $H(z)$ dataset. The parameters $\alpha$ and $\Omega_{0,m}$ are dimensionless quantities. The parameters $H_{0,\mathrm{CCC}}$ and $H_0$ are expressed in units of $\rm km\, s^{-1}\rm Mpc^{-1}$. \label{tab:prior+post}}
    \begin{tabular}{clcl}
    \hline
    Parameter & Prior Range & Prior Type &  Posterior \\
    \hline
    \multicolumn{4}{c}{CCC+TL}\\
    \hline
    $\log_{10}(H_{0,\mathrm{CCC}})$  & [-1.0, 4.0] & Uniform  &  $0.94^{+0.34}_{-0.57}$ \\
    $\alpha$ & [-10000, 60] & Uniform  & $13.03^{+2.07}_{-5.76}$ \\
    \hline
    \multicolumn{4}{c}{$\Lambda$CDM} \\
    \hline
    $H_0$ & [10.0, 100] &  Uniform  &  $68.90^{+4.08}_{-4.13}$ \\
    $\Omega_{0,\mathrm{m}}$  & [0.0, 1.0] &  Uniform & $0.32^{+0.07}_{-0.06}$ \\
    \hline
    \end{tabular}
\end{table}

\section{Results}\label{sec:result}
\begin{figure*}
\centering
 \includegraphics[width=0.45\linewidth]{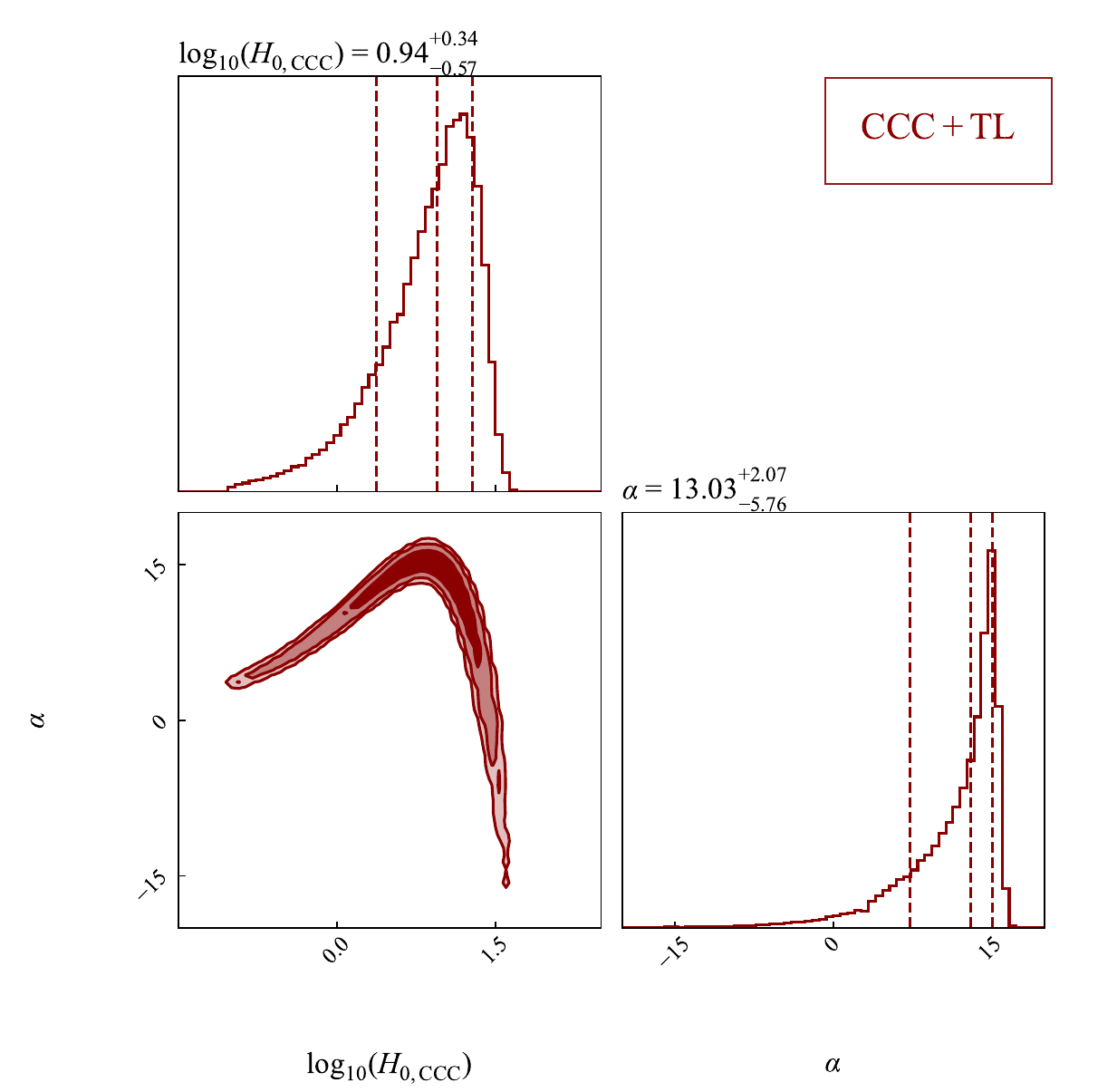}
 \includegraphics[width=0.45\linewidth]{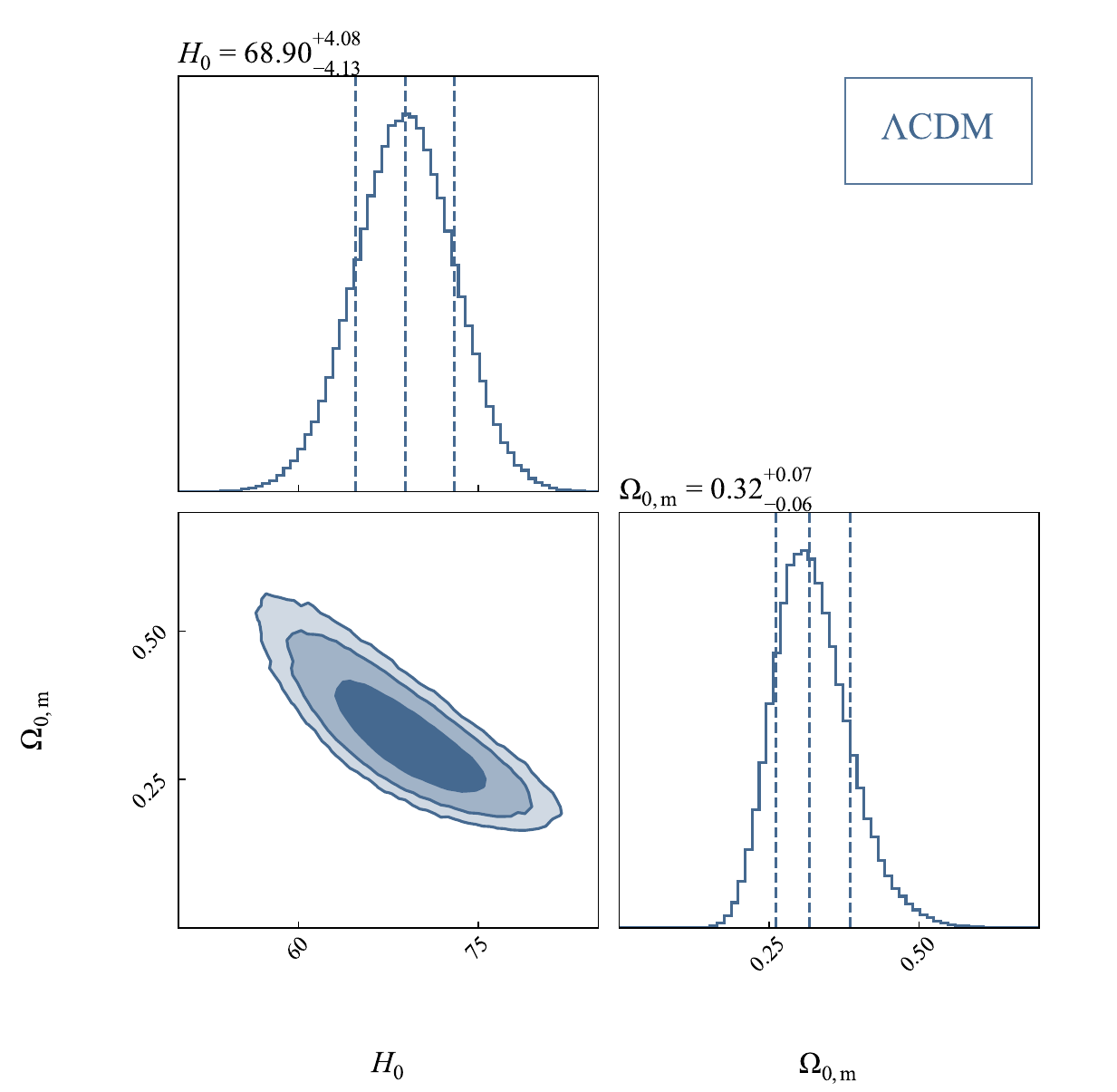}
 \caption{Posterior distributions of parameters for the CCC+TL (Left Panel) and $\Lambda$CDM (Right Panel) models obtained from fitting the $H(z)$ dataset.}
 \label{fig:pdf}
\end{figure*}

After fitting the models to the $H(z)$ data, we obtained the best-fit values and the posterior distributions of the parameters. 
Table \ref{tab:prior+post} summarizes the posteriors (median and $1\sigma$ interval), and Table \ref{tab:result} lists the best-fit parameter combinations and goodness-of-fit statistics. The posterior distributions, shown in Figure \ref{fig:pdf}, deviate significantly from Gaussian for CCC+TL parameters.

Figure \ref{fig:pdf} reveals a significant difference between the $H(z)$-constrained CCC+TL parameters and those derived from the SNe Ia dataset \citep{2023MNRAS.524.3385G}. Most notably, the posterior of speed-of-light variation exponent $\alpha = 13.03^{+2.07}_{-5.76}$ from $H(z)$ strongly contradicts the SNe Ia best-fit value $\alpha = -47.59\pm 0.83$. The SNe Ia $\alpha$ value lies 
far away from the median of the $H(z)$ posterior distribution for $\alpha$, indicating severe tension. This demonstrates that the CCC+TL model cannot adequately describe both the SNe Ia and $H(z)$ datasets simultaneously.

In contrast, for $\Lambda$CDM, the parameters best-fitting the SNe Ia dataset (i.e., $H_0 = 72.99\, \rm km\, s^{-1}\, Mpc^{-1}$, and $\Omega_{0,\mathrm{m}} = 0.35$; \cite{2023MNRAS.524.3385G}) fall comfortably within the $1\sigma$ region of the posterior distribution derived from the $H(z)$ data (i.e., $H_0 = 68.90^{+4.08}_{-4.13}\, \rm km\, s^{-1}\, Mpc^{-1}$, and $\Omega_{0,\mathrm{m}}= 0.32^{+0.07}_{-0.06}$; see Table~\ref{tab:prior+post} and Figure~\ref{fig:pdf}), showing consistency.

We observed that the posterior distributions of the CCC+TL model parameters, as shown in Figure \ref{fig:pdf}, deviate from a Gaussian distribution. To ensure the robustness of these results, we rigorously examined the sampling efficiency and convergence of our MCMC chains. We sampled the posterior distributions using the \textit{emcee} package \citep{2013PASP..125..306F} with 32 walkers running for 20,000 steps, achieving a chain length exceeding 70 times the integrated autocorrelation time ($\tau \approx 258$). The acceptance fraction was $\approx 0.40$, indicating efficient sampling. Throughout this work, the quoted 'best-fit' parameters refer to the maximum likelihood values, whereas the constraints reported in Table \ref{tab:prior+post} correspond to the median values and $1\sigma$ credible intervals from the posterior distributions.

Figure \ref{fig:data} compares the $H(z)$ predictions for the best-fit SNe Ia parameters of both models with the actual $H(z)$ data and the best fits obtained directly from $H(z)$. The $H(z)$ curve corresponding to the CCC+TL SNe Ia parameters (red dotted line) deviates markedly from the measured values, especially at higher redshifts. The $\Lambda$CDM model (blue line), however, fits both the SNe Ia and $H(z)$ datasets well with consistent parameters.

To quantitatively compare the models' performance on the $H(z)$ data, we compute the $\chi^2$ and maximum log-likelihood ($\ln{\mathcal{L}_{\mathrm{max}}}$) for different model configurations (Table \ref{tab:result}). Differences relative to $\Lambda$CDM (i.e., $\Delta \chi^2$) are also presented. To compare the performance of different models, we utilize the Akaike Information Criterion (AIC) \citep{1974ITAC...19..716A} and the Bayesian Information Criterion (BIC) \citep{1978AnSta...6..461S}, defined as:
\begin{align}
    \mathrm{AIC} &= 2k - 2\ln \mathcal{L}_{\mathrm{max}}, \\
    \mathrm{BIC} &= k\ln N - 2\ln \mathcal{L}_{\mathrm{max}},
\end{align}
where $k$ is the number of free parameters, $N$ is the number of data points, and $\mathcal{L}_{\mathrm{max}}$ is the maximum likelihood value. Both criteria penalize model complexity, with BIC imposing a stricter penalty for the number of parameters given our sample size ($N=32$). A lower value of AIC or BIC indicates a preferred model. The results are summarized in Table \ref{tab:result}.

\begin{table}
\centering
\caption{Comparison of goodness-of-fit metrics for the $H(z)$ dataset ($N=32$). $\Delta$ values are defined relative to the $\Lambda\mathrm{CDM}$ model. Note that for models with the same number of parameters ($k=2$), $\Delta\mathrm{AIC} \equiv \Delta\mathrm{BIC} \equiv \Delta\chi^2$. }
\label{tab:result}
\setlength{\tabcolsep}{4pt} 
\begin{tabular}{l|c|c|c|c|c|c}
\hline
\textrm{Model} & \textrm{$\Theta^a$} & \textrm{$\chi^2$} & \textrm{$\ln{\mathcal{L}_{\mathrm{max}}}$} & \textrm{AIC} & \textrm{BIC} & \textrm{$\Delta\chi^2$} \\
\hline
CCC+TL: SNe Ia fit & A & 76.02 & -159.85 & 323.70 & 326.63 & 61.52 \\
CCC+TL: $H(z)$ free fit & B & 17.21 & -130.45 & 264.90 & 267.83 & 2.71 \\
$\Lambda$CDM: $H(z)$ free fit & C & 14.50 & -129.10 & 262.20 & 265.13 & 0.00 \\
\hline
\end{tabular}
\\
\begin{flushleft} 
\small
Note $^a$: Parameters $\Theta$:\\
A: $H_{0,\mathrm{CCC}}=59.51, \alpha=-47.59$ fixed from SNe Ia \citep{2023MNRAS.524.3385G}. \\
B: $H_{0,\mathrm{CCC}}=8.73, \alpha=15.11$ (best-fit from this work). \\
C: $H_0=69.43, \Omega_{0,\mathrm{m}}= 0.31$ (best-fit from this work).
\end{flushleft}
\end{table}

Table \ref{tab:result} demonstrates that $\Lambda$CDM provides the best fit to the $H(z)$ data, yielding the highest likelihood (least negative $\ln{\mathcal{L}_{\mathrm{max}}}$). The $\Delta \chi^2 = \Delta {\rm AIC}= \Delta {\rm BIC} =61.52$ between the SNe Ia-optimized CCC+TL model and $\Lambda$CDM illustrates strong tension. Furthermore, even the freely fitted CCC+TL model to $H(z)$ yields larger $\chi^2$,$\rm AIC$ or $\rm BIC$ values than $\Lambda$CDM, indicating that $\Lambda$CDM describes the $H(z)$ data better.

Our findings reveal significant tensions for the CCC+TL model:
1.  The parameter set derived from the SNe Ia data fails to describe the $H(z)$ measurements from cosmic chronometers accurately. In stark contrast, the standard $\Lambda$CDM model exhibits no such conflict.
2.  A rigorous statistical comparison using both $\Delta\chi$, $\Delta \rm AIC$ and $\Delta \rm BIC$ strongly favors $\Lambda$CDM over CCC+TL in describing the $H(z)$ data.  This also indicates a significant tension between the model parameters constrained by SNe Ia and the observed expansion history.
3. Specifically, fixing $\alpha$ to the SN Ia best-fit value ($\alpha = -47.59$) yields a $\Delta \chi^2 = 58.81$ relative to the free-$\alpha$ fit for the $H(z)$ data, corresponding to a likelihood ratio of $\mathcal{R} = e^{-\Delta \chi^2/2}  \approx 1.7 \times 10^{-14}$. This robust, likelihood-based assessment confirms that the two datasets are mutually exclusive within the CCC+TL framework at a high confidence level. This indicates a fundamental inconsistency within the CCC+TL framework itself – the parameters required to fit the distance modulus (SNe Ia) are incompatible with those required to fit the expansion rate $H(z)$.

\section{Conclusion and Discussion}\label{sec:dis}
These results demonstrate that while the CCC+TL model may offer an explanation for JWST's angular diameter observations of high-$z$ galaxies, it faces severe challenges when confronted with the model-independent record of the universe's expansion history encoded in cosmic chronometers. The significant tension with $H(z)$ data and the internal inconsistency revealed between SNe Ia and $H(z)$ constraints pose substantial difficulties for the CCC+TL model's overall viability. $\Lambda$CDM, though facing challenges from JWST's early galaxies, is well consistent with the totality of the expansion history data. This work underscores the critical importance of multi-probe consistency tests, particularly utilizing model-independent measures like $H(z)$ from cosmic chronometers, for evaluating novel cosmological frameworks proposed in the exciting, yet challenging, new era opened by JWST. The resolution to the "JWST high-$z$ galaxy size tension" likely lies either within refined astrophysics operating within $\Lambda$CDM (e.g., galaxy size evolution) or more exotically a new  cosmology.

The primary motivation for proposing CCC+TL \citep{2023MNRAS.524.3385G} was the unexpectedly small apparent angular diameters ($\theta$) of high-redshift ($z \sim 10$) galaxies observed by JWST. However, this phenomenon may have a natural explanation within $\Lambda$CDM through intrinsic galaxy size evolution in the early universe. For instance, recent JWST observations indicate strong evolution in the intrinsic sizes of early galaxies \citep{2023ApJ...955L..12B,2025ApJ...988..196M,2025ApJ...991..222O,2025AstBu..80..337R}. Furthermore, studies have identified galaxies with remarkably compact luminous regions \citep{2023Sci...380..416W}, suggesting that the intrinsic properties and evolution of early galaxies require deeper investigation.

Beyond the Hubble parameter constraints presented here, there are other independent cosmological probes, such as HII regions \citep{2025MNRAS.538.1264C}, gamma-ray bursts (GRBs) \citep{2015NewAR..67....1W,2025ApJ...988L..71W}, and fast radio bursts (FRBs) \citep{2014PhRvD..89j7303Z,2022MNRAS.515L...1W,2025ApJ...981....9W}. A robust cosmological model must withstand scrutiny from multiple such probes. Though $\Lambda$CDM faces tensions (e.g., $H_0$, $S_8$), it remains relatively robust across this broader range of tests.

\begin{figure}[htbp!]
\centering
 \includegraphics[width=\linewidth]{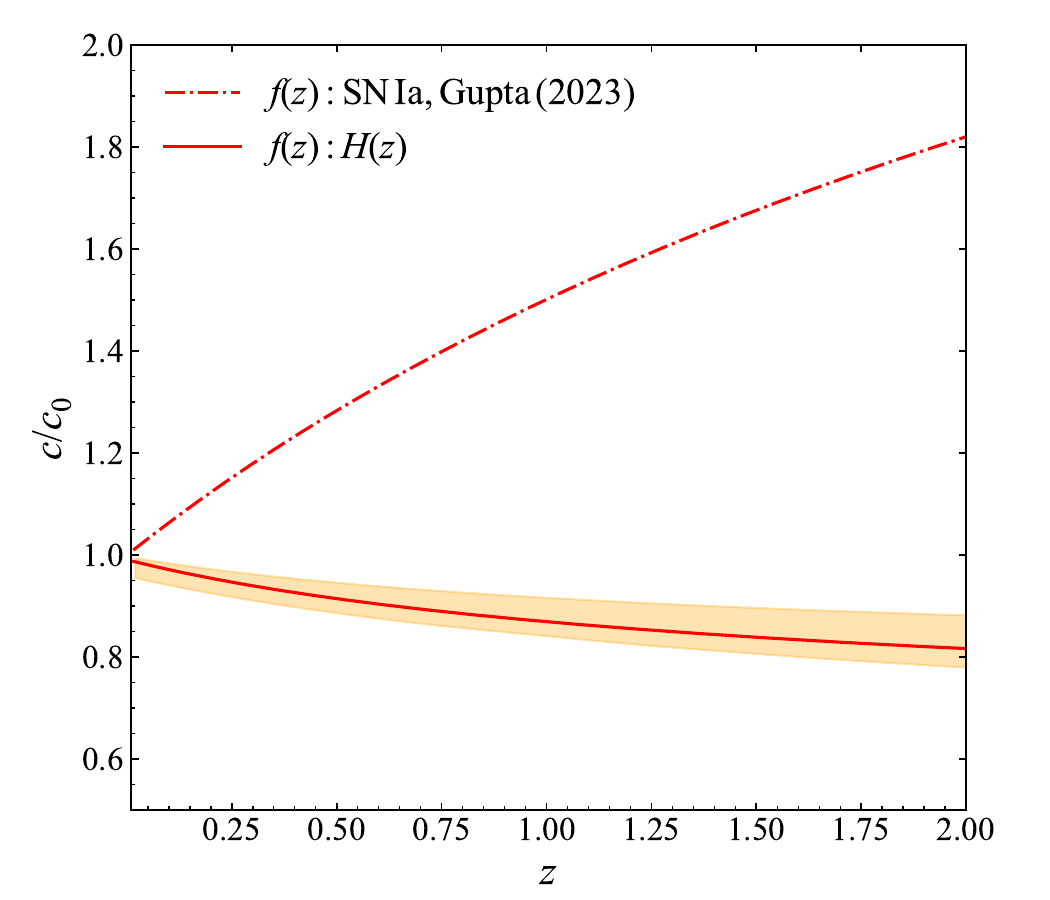}
 \caption{The variation function of the speed of light ($f(z)=c/c_0$) within the CCC+TL model. The red dash-dotted and solid lines correspond to the best-fit results for the SNe Ia \citep{2023MNRAS.524.3385G} and $H(z)$ datasets, respectively. The orange region is the 68\% range fitted with $H(z)$ dataset.  } 
 \label{fig:f_z&c/c0}
\end{figure}

Within the CCC+TL framework, the cosmic expansion history is governed by the function $f(t) = e^{\alpha (t - t_0)}$, describing the exponential variation of the speed of light with time, coupled with a covariation of the gravitational constant $G \propto c^3$. The solution parameterized in Eq. (\ref{eq:f_z}) determines $f(z_{\mathrm{CCC}})$. Figure \ref{fig:f_z&c/c0} shows the speed of light variation function $c/c_0 = f$ (since $c \propto f$ in the model \citep{2023MNRAS.524.3385G}) corresponding to the best-fit SNe Ia parameters (red dotted line) and $H(z)$ parameters (red solid line) of the CCC+TL model.

As shown by the red dash-dotted line in Figure \ref{fig:f_z&c/c0}, the CCC+TL model parameters fitted to the cosmological luminosity distance modulus dataset (SNe Ia) favor a rapidly decreasing speed of light with cosmic age. Conversely, the parameter results from our $H(z)$ fitting suggest minimal evolution of the speed of light (red solid line in Figure \ref{fig:f_z&c/c0}). This fundamental tension stems from the opposite signs of the best-fit $\alpha$ parameter in SNe Ia versus $H(z)$ datasets. 

Recent studies have constrained cosmological evolution of fundamental constants like $G$ \citep{2025A&A...697A.109L} and $c$ \citep{2024MNRAS.527.7713M,2024JCAP...11..062S,2025PDU....4801947L,2024PDU....4301380M,2014PhRvD..90f3526Q,2025PhLB..86839756H}. 
Furthermore, we find that the speed-of-light variation function $f$ corresponding to our $H(z)$-constrained CCC+TL parameters remains consistent with the model-independent constraints on cosmological $c$-evolution from \citet{2024MNRAS.527.7713M} (see their Figure 3).

In conclusion, our analysis suggests that the tension posed by JWST observations of compact high-$z$ galaxies may not originate from the cosmological model itself, but rather reflect our incomplete understanding of the intrinsic properties and evolution of galaxies in the early universe.

\section*{Acknowledgements}
We thank the referees for their helpful comments. We thank Prof. Rajendra Gupta and Dr. Utkarsh Kumar for their helpful discussions.
This work is supported by the National Natural Science Foundation of China (No. 12233011), and the Postdoctoral Fellowship Program of CPSF (No. GZB20250738).  G.Y. acknowledge support from the University of Trento and the Provincia Autonoma di Trento through the UniTrento Internal Call for
Research 2023 grant “Searching for Dark Energy off the beaten track” (DARKTRACK, grant agreement no.E63C22000500003).

\section*{Data Availability}

We adopt 32 independent $H(z)$ measurements \citep{2014RAA....14.1221Z,2005PhRvD..71l3001S,2012JCAP...08..006M,2015MNRAS.450L..16M,2016JCAP...05..014M,2010JCAP...02..008S,2017MNRAS.467.3239R}, which are organized at \url{https://apps.difa.unibo.it/files/people/Str957-cluster/astro/CC_data/data_CC.dat}.



\bibliographystyle{mnras}
\bibliography{biblio} 








\bsp	
\label{lastpage}
\end{CJK*}
\end{document}